 \documentclass[aps,pre,twocolumn,showpacs]{revtex4}

\usepackage{graphicx}
\usepackage{amssymb}
\usepackage{amsmath}

\begin{document}

\title{Scaling in self-organized criticality
from interface depinning?}
     
\author{Mikko Alava\footnote{email:mja@fyslab.hut.fi}} 
\affiliation{ 
Helsinki University of Technology, Laboratory of Physics,  
HUT-02105 Finland
} 
\date{\today}

\begin{abstract} 
The avalanche properties of models that exhibit
'self-organized criticality' (SOC) are still mostly
awaiting theoretical explanations. A recent mapping
(Europhys. Lett.~53, 569) of many sandpile models
to interface depinning is presented first,
to understand how to reach the SOC ensemble and the differences of this
ensemble with the 
usual depinning scenario. In order to derive the SOC avalanche 
exponents from those of the depinning critical point, a geometric
description is discussed, of the quenched landscape in which the 'interface'
measuring the integrated activity moves. It turns out that there 
are two main alternatives concerning the scaling properties
of the SOC ensemble. These are outlined
in one dimension in the light of scaling arguments and 
numerical simulations of a sandpile model which is in the quenched
Edwards-Wilkinson universality class.

\end{abstract} 

\maketitle   
 
\date{\today} 
 
\section{Introduction} 
The concept of reaching ``self-organized'' criticality (SOC) 
in a model without any apparent tuning parameter draws still
attraction \cite{AIP}. In many real-life systems power-law probability
distributions are met, and it is thus an obvious question to
ask why they would resemble ordinary critical phenomena.
To this end, sandpile models have been the prevalent theoretical
playground for the last fifteen years. Only recently their
understanding has finally started to take shape. Two supporting
approaches have been developed. The crucial notion is that
the critical state in these models draws from both the boundary
conditions and the drive, and also has a generic, field-theoretical
description. This can be formulated as a variant of the directed
percolation-style Reggeon field theory (RFT) \cite{guys},
or as a mapping to {\em interface depinning} \cite{ala1,ala2,ala3,previous}.
The gist of this mapping is based
on the description of the {\em history} of the sandpile model
and its dynamics via a stochastic differential equation, the
quenched Edwards-Wilkinson (qEW) equation \cite{qew1,qew2}. 
Likewise a suitable RFT
for sandpiles includes by necessity a density conserving term that
accounts for the effects of diffusional transport of particles or
grains.

In this article the description of sandpiles through interface
depinning is reviewed. The issue of particular interest to us
is the physics of driven interfaces that describe sandpile models.
To this end Section II introduces in a short fashion the mapping,
and discusses the ensemble in which SOC is reached. It is seen
that this is not any of the normal ones familiar in the context
of the qEW, say. The next section is devoted to a discussion
of the standard properties of interface scaling at the depinning
transition, and, correspondingly, the scaling laws usually
formulated for sandpiles depending on the ensemble. We next
concentrate in particular on a {\em geometric description} of depinning.
This is the essential issue in obtaining the scaling
exponents of SOC avalanches: the extension of the theory
of 'elastic pinning paths' to the SOC case. In section IV
numerical results are considered for the one-dimensional
qEW sandpile in the SOC ensemble. This is the {\em simplest}
model system in which one can try to extract a correspondence
with the avalanche and the depinning pictures. This is since the 1D
pinning paths can be discussed without the introduction of extra, 
independent exponents. Finally, section V finishes the paper
with a summary about SOC properties based on recent advances
including the numerical work presented here, and about remaining
open problems.

\section{Mapping sandpiles to interfaces}
Consider a sandpile model with each site $x$ of a
hyper-cubic lattice having $z(x,t)$ grains.
When $z(x,t)$ exceeds a critical threshold
$z_c(x)$, the site is active and topples so (y is a nearest
neighbour of $x$)
\begin{equation}
        z(x,t+1) = z(x,t) - 2d, ~
        z(y,t+1) = z(y,t) + 1.
\end{equation}
$z_c(x)$ is taken to be a random variable, chosen
from a probability distribution $P(z_c)$ again {\em after
each toppling}. If there are no active sites in
the system, one grain is added to a randomly chosen site, 
$z(x,t) \to z(x,t) + 1$.
This is the SOC ensemble, defined
via the drive and the open boundaries, 
characterized by avalanches that may ensue after a grain
has been deposited.
The dynamics of this model can be reinterpreted
through an 'interface', or history $H(x,t)$ which follows the
memory of all the activity at $x$. $H$
counts topplings at site $x$ up to time $t$,
and has the dynamics
\begin{equation}
        H(x,t+1) = \left\{ \begin{array}{ll}
                        H(x,t) + 1,  &  f(x,t) >0,  \\
                        H(x,t),      &  f(x,t) \le 0,
                   \end{array} \right.
                           \label{eq:H(x,t+1)}
\end{equation}
This can be written as a
discrete interface equation 
\begin{equation}
	\frac{\Delta H}{\Delta t}=\theta \left(f(x,t) \right) , 
                           \label{eq:H-def}
\end{equation}
with $\theta(f)$ the step-function forcing it so that
the interface does not move backwards.
The 'local force' is 
\begin{equation}
f(x,t) = z(x,t)-z_{c}(x) = n_x^{in} - n_x^{out} - z_{c}(x), 
\end{equation}
in terms of $n_x^{in}$ (grains added to site $x$ up to time $t$) and
$n_x^{out}$ (grains removed from $x$). 
The fluxes $n_{x}^{in}$ and $n_{x}^{out}$ 
can be worked out in terms of the local height
field $H(x,t)$, and a columnar force term $F(x,t)$ 
which counts the number of grains added to site $x$
by the external drive,
\begin{equation}
f(x,t) = \nabla^2 H + F(x,t) - z_c(x,H).
\end{equation}

The step-function, $\theta(f)$, in Equation~(\ref{eq:H-def}),
the condition that the interface  does not move backwards, introduces
an extra noise term $\sigma(x,H)$ - the velocity of the interface is either
one or zero - but for the current example this should not be
relevant. Combining the three sources of effective noise,
$F$, $z_c$ and $\sigma$, one ends at the discretized interface equation 
\begin{equation}
	\frac{\Delta H}{\Delta t}  = 
	        \nabla^2 H + \eta(x,H) + F(x,t) +\sigma(x,H) .
					\label{eq:btw-equ}
\end{equation}
Here the quenched noise $\eta(x,H) = - z_c(x,H)$, and
we have obtained the central difference
discretization of a continuum diffusion equation
with quenched noise, called the linear interface
model (LIM) or the quenched Edwards-Wilkinson equation 
\cite{qew1,qew2}.
The continuum limit reads 
$\partial_t h(x,t) = \nu \nabla^2 h (x,t) + F + \eta(x,h(x,t))$.

\begin{figure}  
\includegraphics[width=7cm
]{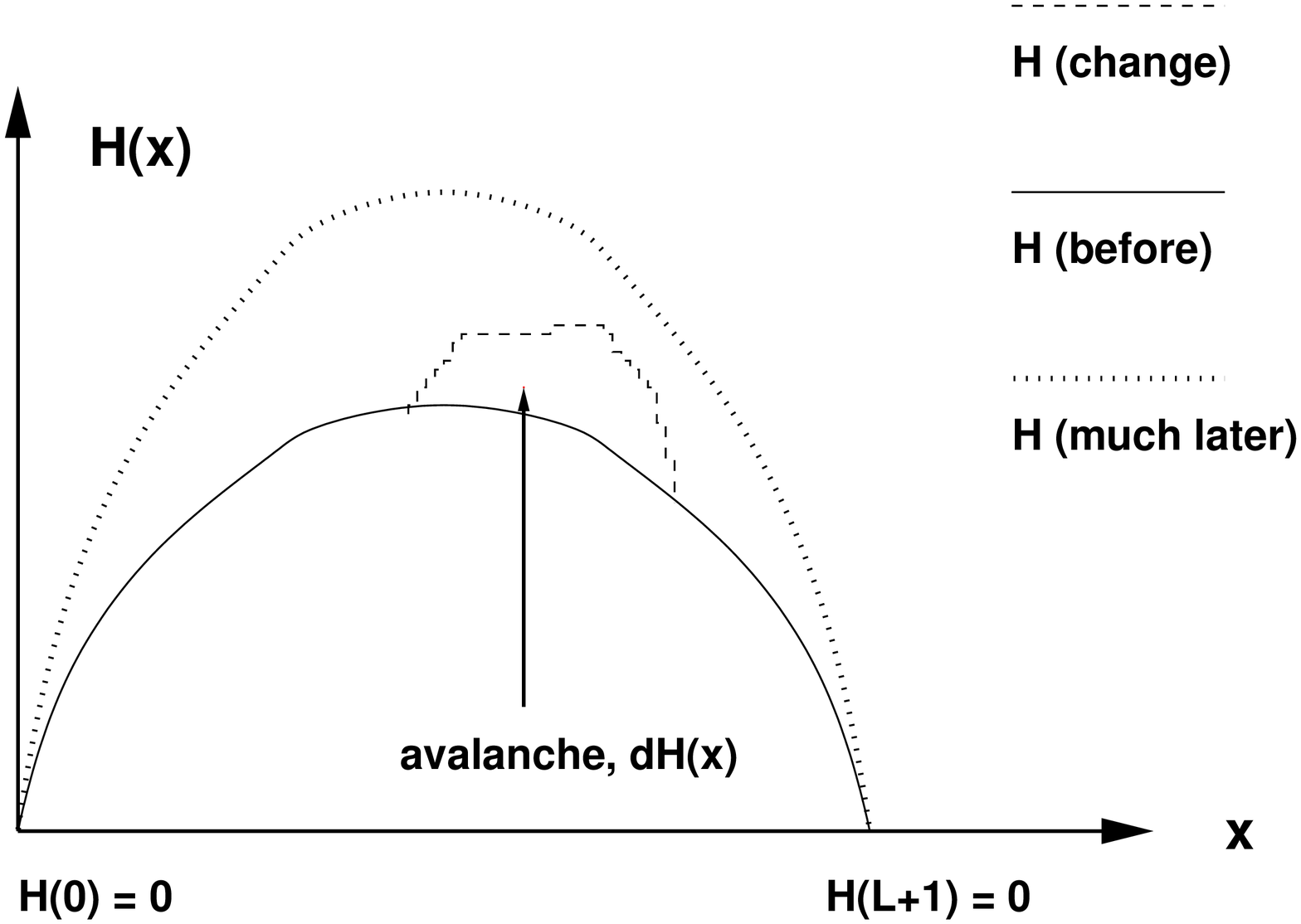}
\caption{One-dimensional schematic example of
the interface or history representation of a
SOC sandpile model. The mean-field interface follows an
average parabolic shape, which also implies that
in the SOC steady-state $v(x)$ is parabolic.
Notice the boundary conditions $H=0$ that
ensure the loss of particles (equalling the
increased elastic energy) in that state.}
\label{ense}
\end{figure}

The SOC ensemble is illustrated in Figure (\ref{ense}).
Once the local force is increased, by adding a grain and
making $F(x,t) \rightarrow F(x,t)+1$ at the chosen $x$ an
avalanche starts since the force overcomes the pinning force,
$\nabla^2 H + \eta(x,H) + F(x,t) > 0$. The interface moves
at $x$ by one step, $\Delta H(x)=1$. In the subsequent dynamics
of the avalanche the columnar force term $F(x)$ does not change.
For the SOC sandpiles, the correct choice of the interface boundary
condition is $H = 0$ which is to be imposed at two ``extra sites''
($x=0$, $x=L+1$ for a system of size $L$ in 1D). The 
ever-increasing $\langle F(t) \rangle$ leads to an average
parabolic interface shape via the cancellation of the Laplacian
by $F$. It is to be stressed that this implicitly contains
the physics of the SOC state: the driving force $F(x,t)$ is increased
so slowly that the avalanches do not overlap and are therefore
well-defined \cite{ala1}.

\section{Scaling properties of ensembles}

Equation (\ref{eq:btw-equ}) exhibits a {\em depinning transition} at 
a threshold force  $F_c$ in a ¨normal¨ ensemble.
The interface configuration and dynamics develop 
critical correlations in the vicinity of the critical point. 
For the case of point-correlated disorder the normal way
to analyze the LIM is to use the functional renormalization 
group method. One-loop expressions for the exponents are found
in papers by Nattermann et al.,
Narayan and Fisher \cite{qew1,qew2}, and in the more recent ones
by Le Doussal and collaborators \cite{Doussal}.

The problem is technically and from the fundamental viewpoint
difficult, since the whole disorder correlator is renormalized.
The point-random field noise term forms one of the universality
classes in this problem \cite{qew1,qew2}; others are 
the LIM  with columnar noise \cite{columnar}
and the quenched Kardar-Parisi-Zhang equation \cite{TKD,HHZ,Barabasi,Gabor}
in more general terms.
The mapping of SOC models to variants of LIM holds however
some surprises. For instance, while the conjecture that
the Manna model \cite{manna} should be in the usual point-disorder LIM 
class seems to be true in 2D, this is clearly not so in 1D
if considered in the depinning ensemble \cite{mannarefs}. 
The complications arise since an arbitrary choice of sandpile rules can 
lead to {\em non-standard noise correlations} that
do not need to have a priori the same RG fixed point as the random field
or force case.

In the interface language the relevant exponents to describe
the depinning phase transition are: $\nu$ (the correlation
length exponent), $z$ (the dynamical exponent), and $\chi$,
the roughness exponent. Moreover assuming that the avalanche
dynamics suffices to describe the interface dynamics off the
critical point, $\theta=\nu(z-\chi)$ holds for the velocity
exponent, with $v \sim (F - F_c)^\theta$
\cite{qew2,qew1}.
For point-like disorder the first-loop functional RG results
cited above read $\chi = (4-d)/3$, and $z = 2 - (4-d)/9$.
Notice the exponent relation $\nu = 1/(2-\chi)$ which 
manifests together with the $\theta$-exponent relation
the fact that there is only one temporal and one spatial
scale at the critical point. 

The typical quantity to be measured in the interface
context is the interface width $w$ (mean fluctuation) which
in most cases equals other measures like
the two-point correlation functions and the structure
factor \cite{Barabasi}. From the sandpile viewpoint, these measure the
correlations and fluctuations in the activity history, that is
in the avalanche series. Initially the width grows, as 
a function of time, as $w \sim t^\beta$ defining the 
'growth exponent' $\beta$ until either saturation
is reached (with a finite order parameter, $v$), or
the interface gets pinned at or below $F_c$. Scaling now implies
$\beta z = \chi$, as for general interface models.

The LIM obeys an important invariant,
with the static response scaling as \cite{qew2}
$\chi(q,\omega=0) \sim q^{-2}$, so that
\begin{equation}
	\gamma/\nu = 2   .
		\label{eq:gamma/nu}
\end{equation}
For forces below $F_c$, the (bulk) response of the interface triggered
by a small increase in $F$ is
$	\chi_{\rm bulk} \equiv {d \left< H \right>} / {dF} 
					\sim (F_c-F)^{-\gamma}$.
By assuming that the avalanche due to a point seed scales as
$\Delta H \sim s \sim \ell^D$, $D={d+\chi}$
(since $H$ scales with the roughness as $\ell^\chi$), 
a hyperscaling relation can be derived for $\gamma$.
Right at the critical point 
\cite{qew2,qew1},
the roughness of the interface scales as $\ell^\chi$ and assuming that
$\Delta \left< H \right>$ will scale in the same way it follows 
\begin{equation}
	\gamma = 1 + \chi \nu .
		\label{eq:gamma}
\end{equation}
This also implies $\chi + 1/\nu = 2$, as noted above.
The standard scaling relations are valid for parallel dynamics: 
all sites with $\partial H / \partial t > 0$ are updated in parallel. 
For extremal drive criticality (updating one unstable site at a time)
the dynamic exponent reads  $z_{\rm ED}=2$.

The 1D LIM is a bit more peculiar than one might expect. Numerics (which 
has recently been matched by the 2-loop RG results of LeDoussal
et al.) implies that $\chi \sim 1.2 \dots 1.25 $ which is
larger than the 1-loop and scaling argument result $\chi_{1d} =1$.
The physical interpretation for the fact that
$\chi > 1$ has been dubbed
``anomalous scaling'' \cite{Lesch,Juanma}, and arises
from a divergent mean height difference between neighboring 
sites, with $t \to \infty$.

The SOC steady-state is characterized by the probability to have 
avalanches of lifetime $t$
and size $s$ which follow power-law distributions:
$
	p(t) = t^{-\tau_t} f_t(t/L^z)
$
and
$
	p(s) = s^{-\tau} f(s/L^D)  ,
$
with $s\sim t^{D/z}$ and $z(\tau_t-1)=D(\tau-1)$.
One can also characterize the avalanche by its linear dimension,
	$p(\ell) = \ell^{-\tau_\ell} f_\ell(\ell/L)$,
with $\tau_\ell = 1+D(\tau -1)$.
Here the size scales as $s\sim \ell^D$
and the (spatial) area as $\ell^{d}$ (for compact avalanches)
with $\ell$ the linear dimension.
The fact that each added grain will perform of the order of $L^2$ 
\cite{Dhar}
topplings before leaving the system leads to the fundamental result
\begin{equation}
        \left< s \right> \sim L^2
			\label{eq:<s>}
\end{equation}
independent of dimension \cite{tang:1988}.
Thus, $\tau=2-2/D$ and $\tau_t = 1 + (D-2)/z$.
Equation~(\ref{eq:<s>})
yields $\gamma/\nu=2$, where $\gamma$ describes the divergence 
of the susceptibility 
(bulk response to a bulk field) near a critical point,
if one generalizes the exponent relation of the
depinning ensemble to the SOC case. In the particular
case of an increasing drive and a bulk dissipation, which 
induces a term $-kH$ to the qEW equation, this should, evidently, 
be valid \cite{bulk}.

\section{The one-dimensional QEW sandpile: geometric
description of avalanches}
To go beyond the scaling exponents to a description
of the {\em probability distributions} at the critical
point of any particular ensemble is a more challenging
task. This is easiest in one dimension, which we 
thus discuss here. The most well-known case in which
the geometry of the random quenched landscape allows
to use a self-affine picture of the progress of an
interface through it is given by directed percolation
depinning (DPD), the quenched KPZ cousin of the qEW
\cite{TKD,HHZ,Barabasi}.

In this language the interface moves, e.g. in the 
case of an applied extremal drive, via a succession of 
¨punctuation events¨. The interface is assumed to 
invade the voids of a connected network in each
of these events, in 'bursts'.
In between these events, the pinned interface
is mapped into a connected path on the backbone of
a suitable (elastic) percolation problem \cite{pinningpaths},
For DPD the analogy  is more or less clear, and for the qEW one speaks of {\em
elastic pinning paths}. Such paths have the characteristics
that the RHS of the qEW is always negative semi-definite,
ie. $f \leq 0$. It is not known rigorously whether such
paths at criticality follow strictly the DPD -like scaling
properties.
There are two fundamental issues: the geometric properties
of avalanches (the scaling of voids,
or whether the relation of size vs. area 
can be characterized with a ¨local¨ roughness exponent
$\chi_{loc}$) and the probability to induce an avalanche,
or ¨punctuation event¨ if the interface is pushed at any
particular spot (this issue is still open to discussion,
see \cite{huber,zeitak,jost}).

Assuming that the DPD analogy works \cite{huber}, it 
follows in the
depinning ensemble for the avalanche size
exponent 
\begin{equation}
\tau_{s,dep} = 1 + (1/(1+\chi_{loc}))(1-1/\nu)
\end{equation}
which using eg. the exponent relation $\nu = 1/(2-\chi)$ 
and a reasonable value for $\chi_{loc}$, taken to equal
the global $\chi$, produces
\begin{equation}
\tau_{s,dep} \approx 1.08.
\end{equation}

For the SOC ensemble, the description of the critical state
is given in terms of the avalanche distribution exponents
- note that due to the inhomogeneity of the ensemble e.g. perturbing
the system off the critical point is more complicated \cite{ala3}.
One has then to ask the question
what is the prediction of pinning-path picture. Due to the
symmetries (static response) of the qEW it could be assumed that the typically
parabolic interface profile is irrelevant, except perhaps
for finite size corrections induced by the boundary condition
that is imposed on the pinning paths since $H=0$ at the 
edges. This would imply, in particular, that for the
SOC avalanche exponent it is found that
\begin{equation}
\tau_{s,SOC} \rightarrow \tau_{s,dep}.
\end{equation}
A parallel approach is to use the invariant (\ref{eq:<s>})
which leads to the prediction
\begin{equation}
\tau_{s,SOC} =2-2/D = 2-2/(1+\chi_{loc})
\end{equation}
by using the roughness exponent to estimate
the cut-off dimension of avalanches.
This is identical with the pinning path estimate above.

These are then the major issues:
is the depinning geometric description applicable
to the SOC ensemble and if not why? The answer
should depend only on fundamental similarities or
differences between the ensembles, and not on the
particular model - class of qEW-like models - nor
the dimensionality. The study of the outcome is
the easiest in 1D whereas in higher dimensions 
further independent exponents are needed, i.e. assumptions
to describe the probability distributions since the
avalanches have in addition to an area vs. volume
relation also a perimeter length vs. area
one \cite{higherdim,zeitak}.

To study the issue we next outline some numerical
results on the 1D qEW/LIM, obtained from a Leschhorn-like
cellular automaton \cite{Lesch,jost} for the interface problem. 
The system is run as a sandpile in that after the 
interface gets pinned a new avalanche initiation
is done by increasing the local force $F_i$ at a
random location by one. System sizes upto $L=8192$
have been studied, with 2 $\times$ 10$^6$ avalanches for
the largest sizes \cite{tobe}. The resulting
avalanche size distributions, after logarithmic
binning, imply that
the effective $\tau$-exponent varies with $L$, and
that the weight of the power-law -like tail increases
with the size. Meanwhile, the effective roughness
exponent slightly {\em decreases}. Due to the systematic
finite size corrections it does not make sense to try
a normal datacollapse using a fixed $D$ and $\tau_{SOC}$
for all $L$. This is a little bit surprising given that
relation (\ref{eq:<s>}) is fulfilled within numerical
accuracy by the data, and that higher momenta of the
size distributions indicate just simple scaling of the
avalanche size distribution. Certainly more numerical
analysis is called for, but two facts are worth pointing
out. {\em First}, even for the largest system size the
effective $\tau$ is way off from the ¨predicted¨ 1.08
(1.024 for $L=8192$). {\em Second}, by blindly - without
any a priori justification - using the scaling Ansatz
\begin{equation}
\tau_{SOC} (L) = \tau_{SOC} (L = \infty) + \Delta \tau (L),
\end{equation}
with a ¨best-working¨ function $\Delta \tau$, one observes that 
there is an apparent power-law correction to the exponent. This
extrapolates very slowly in $L$ 
to $\tau_{SOC} (\infty) \sim 1.115$, ie. clearly
off the depinning ensemble value. 
From $\tau_{SOC}$ it would be in principle
possible to derive the other exponents as well, 
using the consequences of Eq.~(\ref{eq:<s>}),

\section{Conclusions}
Above, we have discussed a strategy to understand
so-called ¨self-organized criticality¨ by mapping
the history of a sandpile model to a driven interface
in a random medium. For SOC models this idea is useful
since it helps to understand universality classes (via
noise terms generated) though there are theoretical
challenges in the understanding of the possible
classes (RG fixed points), and in the role of the ¨discretization¨ 
(called $\sigma$-noise above). Such work follows the historical
connections of SOC to depinning, and extends it by 
explicitly constructing the right ensemble to reach
SOC, and by outlining a general strategy to understand
various models. Applied to other absorbing state phase
transitions, in general,
the history/interface description should be
of interest. In some cases (e.g. the contact process)
it can give rise to new, seemingly independent exponents \cite{dm}.

Once one has defined the right ensemble for SOC in
interface depinning, the most pertinent question
becomes if the usual avalanche exponents in SOC can
be derived from those of the depinning
transition. Here we have addressed the question, but
since the outcome is still open, it is worth reiterating
the two possibilities. Either eventually, by
studying large enough systems numerically, the
exponents of the statistically homogeneous ensemble
are recovered once
finite size effects become negligible.
Or, it becomes apparent that the SOC ensemble {\em is
an independent one}. The microscopic reason for this
would be that the density of grains (average force for the
interface) is non-uniform: in one dimension it is easy
to see that a site $x$ will get a larger grain flux from
its neighbor on the bulk side and a smaller one from the
boundary side - more trivially, there is a net flux of
grains towards the boundaries. This inhomogeneity may 
persist in the thermodynamic limit, in which case the
avalanche relations will be determined by its scaling
properties. 

\begin{figure}  
\includegraphics[width=7cm
]{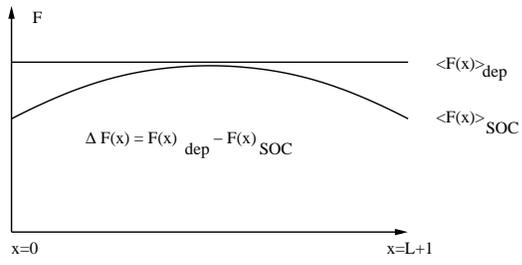}
\caption{The {\em average} force per site, $\langle f(x) \rangle$ in
the depinning and SOC ensembles. $\langle f \rangle$ is to be computed
by computing the number of grains per site between the avanlanches.}
\label{sche}
\end{figure}

Consider the idea depicted in Figure (\ref{sche}).
The integral of the average force deviation at $x$,
$\Delta F(x) = F_{dep} - F(x)_{SOC}$, where $F_{dep}$
and $F(x)_{SOC}$ are averages at the critical points
of the ensembles, 
can be used to define a finite-size correction to
the normal critical point, as  $\int \Delta F(x) dx \equiv
\delta F(L)$. It is seen that $\delta F$ will be dependent
on the exact scaling function of $F(x)_{SOC}$, computing which
is thus the crucial issue. It will give indirectly the
{\em true} correlation length exponent $\nu_{SOC}$ in the SOC ensemble
via the $L$-dependence of the finite size correction, which
may or may not be the same as for the depinning case.
The implication can be rephrased so that the usual 
exponents like $\nu$ are derived from the RG in an ensemble with
statistical translational invariance in the $x$-direction.
It is thus not obvious whether the properties
of self-organized critical state actually 
follow from boundary -induced criticality or are
related to the usual depinning one.
In this respect the usual SOC models discussed
here are inherently more complicated than the boundary
driven cases like the so-called ¨rice-pile¨ model.
This is in particular true if the symmetries of the 
depinning transition are broken by the SOC ensemble,
as is the case for the quenched KPZ equation \cite{Gabor}.

To summarize, the above problem is central
in understanding SOC-like systems. It may
also be tackled, perhaps, from the viewpoint of
absorbing state phase transitions and their field-theoretical
description which provides a parallel route to the
interface one. There are many other interesting issues,
like the early-stage dynamics (growth exponent $\beta$,
suitably defined for the SOC ensemble), the question
of the pinning path/manifolds in higher dimensions 
and so forth. For all these it is invaluable that one
can resort to continuum descriptions of the SOC sandpiles,
that also seem to be inter-connected, in an intriguing
fashion \cite{alepp,am}.

\vspace{1cm}
{\centerline{\bf ACKNOWLEDGMENTS}}
The results presented here have been in part obtained
in an enjoyable collaboration with  K. B. Lauritsen,
M. Mu\'noz, R. Dickman, A. Vespignani, and S. Zapperi.
The Academy of Finland's Center of Excellence Program
is thanked for financial support.

\newpage 

\begin{figure}  
\includegraphics[width=7cm
]{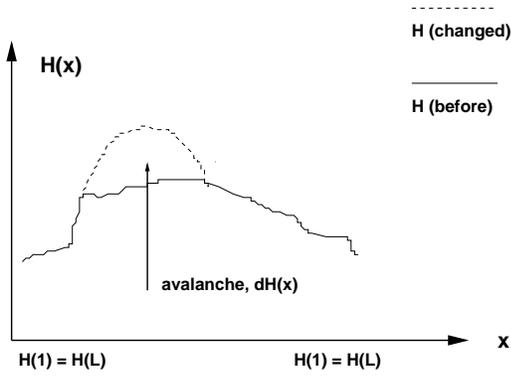}
\caption{An avalanche at the critical point
of the normal, translationally invariant,
¨depinning ensemble¨. The statistical 
description of such avalanches follows
the scaling law $\langle s \rangle
= l^{1+\chi}$, where $l$ is (here in 1D) the
area of the avalanche and $s$ its size.}
\label{aval}
\end{figure}
 
\end{document}